# Information and long term memory in calcium signals

Ildefonso M. De la Fuente[1,2], Iker Malaina[2], Alberto Pérez-Samartín[3], Jesús M. Cortés[4], Asier Erramuzpe[4], María Dolores Boyano[5], Carlos Bringas[1], María Fedetz[1] and Luis Martínez[2].

1. Department of Biochemistry and Pharmacology, Institute of Parasitology and Biomedicine "López-Neyra", CSIC, Granada, Spain.
2. Department of Mathematics, Faculty of Science and Technology, University of the Basque Country, UPV/EHU. Leioa, Spain.
3. Department of Neurosciences, Faculty of Medicine and Dentistry, University of the Basque Country, UPV/EHU. Leioa, Spain.
4. Biocruces Health Research Institute. Hospital Universitario de Cruces. Barakaldo, Spain. Ikerbasque: The basque foundation for Science.
5. Department of Cell Biology and Histology, School Faculty of Medicine and Dentistry, University of the Basque Country, UPV/EHU. Leioa, Spain.

**Corresponding Author:**
Ildefonso Mtz. de la Fuente Mtz.
Instituto de Parasitología y Biomedicina "López-Neyra", CSIC,
Parque Tecnológico de Ciencias de la Salud
Avda. del Conocimiento s/n.
18016 Granada. ESPAÑA.
E-Mail: mtpmadei@ehu.es
Tel.: +34- 958-18-16-21; Fax: +34- 958-18-16-32.

# Abstract

Unicellular organisms are open metabolic systems that need to process information about their external environment in order to survive. In most types of tissues and organisms, cells use calcium signaling to carry information from the extracellular side of the plasma membrane to the different metabolic targets of their internal medium. This information might be encoded in the amplitude, frequency, duration, waveform or timing of the calcium oscillations. Thus, specific information coming from extracellular stimuli can be encoded in the calcium signal and decoded again later in different locations within the cell. Despite its cellular importance, little is known about the quantitative informative properties of the calcium concentration dynamics inside the cell. In order to understand some of these informational properties, we have studied experimental $Ca^{2+}$ series of *Xenopus laevis* oocytes under different external pH stimulus. The data has been analyzed by means of information-theoretic approaches such as Conditional Entropy, Information Retention, and other non-linear dynamics tools such as the power spectra, the Largest Lyapunov exponent and the bridge detrended Scaled Window Variance analysis. We have quantified the biomolecular information flows of the experimental data in bits, and essential aspects of the information contained in the experimental calcium fluxes have been exhaustively analyzed. Our main result shows that inside all the studied intracellular $Ca^{2+}$ flows a highly organized informational structure emerge, which exhibit deterministic chaotic behavior, long term memory and complex oscillations of the uncertainty reduction based on past values. The understanding of the informational properties of calcium signals is one of the key elements to elucidate the physiological functional coupling of the cell and the integrative dynamics of cellular life.

**Author Summary**

Cells need a permanent exchange of energy and matter with the external medium to maintain their self-organized biochemical functionality. As a consequence, metabolic life would not be possible without it having the capacity to adequately respond to the large possible combinations of external variables to which the cell is exposed. These external inputs can be understood in terms of molecular information. Calcium signaling is a ubiquitous mode of biological communication, able to carry information from the extracellular side of the plasma membrane to the different metabolic targets in the internal medium. Intracellular calcium signals show complex oscillatory behavior and play an essential role in cellular metabolism participating in a multiplicity of physiological and pathological functions. Many aspects of the molecular mechanisms involved in the generation of intracellular concentration of calcium dynamics have been relatively well studied, but how information is processed is still unknown. Here, we have exhaustively analyzed experimental calcium time series of *Xenopus laevis* oocytes under different pH stimuli using information-based dynamic tools, and our results show that inside the intracellular $Ca^{2+}$ flows emerges a new kind of dynamical informational structure, in which organizes specific information contained in calcium series.

# Introduction

Calcium ions play a fundamental role in cell physiology, particularly in signal transduction processes. In fact, calcium signaling represents one of the few universal intracellular messengers that are able to carry information from the extracellular side of the plasma membrane to different metabolic targets in the internal medium [1, 2].

From a biochemical point of view, one of the essential characteristics of $Ca^{2+}$ ion fluxes is their capacity to exert regulatory effects on many enzymes [3-6], which allows the modulation of the metabolic activity of the cell. As a consequence, calcium signaling participates in a multiplicity of key functions, such as the regulation of cell differentiation [7], cellular migration [8], cell division [9], ROS signaling [10,11], synaptogenesis [12], apoptosis [13,14], autophagy [15], exocytosis [16], neuronal plasticity [17], cytoskeleton activity [18], cellular growth [19], dendritic development [20], neurotransmitter release [21], and gene expression [22-27].

At a molecular level, the mechanisms involved in the generation of calcium signaling essentially result from a complex interplay among extracellular calcium, permeable channels and the activation/inactivation of intracellular $Ca^{2+}$ pools, mainly located at the endoplasmic or sarcoplasmic reticulum and mitochondria. These processes are highly regulated by complexly interrelated metabolic reactions [28, 29].

In general, external stimuli are often converted into intracellular $Ca^{2+}$ ion fluxes which encode the input information by means of signal transduction processes, i.e., an extracellular stimulus activates a specific receptor located on either the cell surface or the inside, triggering a biochemical chain of molecular processes, which originates different metabolic responses, a large part of them in the form of intracellular calcium oscillations [30].

In many cells, these molecular processes involve G protein-coupled receptors and the phospholipase C enzymes, located in the cell membrane. Specific phospholipase C can hydrolyse phosphatidylinositol 4,5-bisphosphate (PIP2) into two products: inositol 1,4,5-trisphosphate (IP3) and diacylglycerol (DAG), two classical second messengers. The level of external stimulation on the cell determines the degree of activation of the receptor and therefore can be directly linked to the intracellular IP3 concentration. IP3 mainly diffuses to the endoplasmic reticulum, and binds to its specific receptor, which is located in an intracellular calcium channel; thus releases $Ca^{2+}$ ions from the endoplasmic reticulum, triggering oscillations in the cytoplasmic calcium concentration [31]. As a result of these complex metabolic processes the temporal behavior of the intracellular $Ca^{2+}$ levels are highly regulated [32].

In spite of its physiological importance, many aspects of the molecular mechanisms involved in the generation of the intracellular calcium concentration dynamics are still poorly understood.

Intracellular $Ca^{2+}$ fluxes can either be localized or invade the whole cell; and depending on the cell type and the stimulus, $Ca^{2+}$ signals are characterized by a rich variety of oscillating patterns, in which the frequency and amplitude can vary practically infinitely [28,33].

Calcium concentration rhythms represent a genuine manifestation of the dissipative self-organization in the enzymatic activities of the cell. Experimental observations and numerical studies have shown that metabolic processes shape functional structures in which these kinds of molecular rhythms may spontaneously emerge far from thermodynamic equilibrium [34,35].

It is well established that non-equilibrium states can be a source of order in the sense that metabolic irreversible processes may lead to a new type of dynamic state in which the system becomes ordered in space and time. In fact, the non-linearity associated with irreversible enzymatic reactions seems to be an essential mechanism that may allow dissipative metabolic organization far from thermodynamic equilibrium [36,37].

Self-organization is based on the concept of dissipative structures, and its theoretical roots can be traced back to the Nobel Prize Laureate in Chemistry Ilya Prigogine [38].

Two main kinds of metabolic dissipative structures exist in cellular conditions: temporal oscillations and spatial biochemical waves [34]. Both kinds of dissipative structures can be observed in $Ca^{2+}$ dynamics. Thus, when spatial inhomogeneities develop instabilities in the intracellular medium, in addition to temporal calcium oscillation [39], it may lead to the emergence of spatio-temporal dissipative structures which can take the form of propagating calcium concentration waves [40,41]. Phase-coupled NAD(P)H waves and calcium oscillations have also been observed [42].

Since the first observations of the calcium rhythmic phenomena, experimental investigation of its molecular mechanism has been accompanied by numerous computational modeling approaches. The use of these numerical studies on $Ca^{2+}$ oscillations can lead to different important conclusions mainly about the nonlinear feedback processes involved in the spontaneous generation of oscillations [43-50]. For a review on calcium numerical studies, see Dupont et al. [28]. Despite numerous studies on calcium signalling, how information is processed is still insufficiently known.

Different experimental and numerical analysis have shown that the information might be encoded in the amplitude, frequency, duration, waveform or timing of calcium oscillations [51-54]. Thus, specific information coming from extracellular stimuli can be encoded in the calcium signal and decoded again later in different metabolic targets in the cellular internal medium.

These researches on the calcium information have motivated a number of experimental and theoretical studies. For instance, it has been estimated by means of mathematical simulations the information encoded in a $Ca^{2+}$ series upon hormonal stimulation in hepatocytes [55]. Mutual information has also been used to calculate the amount of information transferred over a calcium signaling channel, using a computational model [56]. Long-term correlations have been observed in calcium-activated potassium channels [57-60]; other biochemical processes also present long-term correlation, for example, the intracellular transport pathway of *Chlamydomonas reinhardtii* [61], the NADPH series of mouse liver cells [62] and the mitochondrial membrane potential of cardiomyocytes [63].

Likewise, the information transfer from the calcium signal to a target enzyme under different physiological conditions has been analyzed using transfer entropy [64] in a computational model, a technique widely used to quantify directed biological interactions [65,66].

Here, we have studied experimental calcium series of *Xenopus laevis* oocytes under different external pH stimuli. The data has been exhaustively analyzed by means of information-theoretic approaches such as Conditional Entropy, Information Retention, and other non-linear dynamics tools, such as the power spectra, the Largest Lyapunov exponent and the bridge detrended Scaled Window Variance analysis. We have quantified the biomolecular information flows of the experimental data in bits, and essential aspects of the information contained in experimental calcium fluxes have been exhaustively analyzed. They correspond to the fractional Brownian motion, and our main result shows that in the intracellular $Ca^{2+}$ flows a highly organized complex informational structure emerges which is characterized by exhibiting long-term memory, deterministic chaotic dynamics and oscillations of the uncertainty in respect of the past values.

# Materials and Methods

**1. Calcium oscillations in *Xenopus laevis* oocytes**

Different concentrations of serum are known to promote chaotic oscillations due to alterations of $Ca^{2+}$ concentrations in the cytoplasm, which, as a consequence, evoke Cl- and K+ currents across the oocyte membrane (67, 68) through to the activation of different $Ca^{2+}$ affinity channels (69). Adult *Xenopus laevis* frogs were obtained from Blades Biological (Cowden, Kent, UK). Oocytes at stage V were plucked from the ovaries and defolliculated by collagenase treatment (type 1, Sigma-Aldrich Quimica, S.A., Madrid, Spain) at 80–630 units/ml in frog Ringer's solution (115 mM NaCl, 2 mM KCl, 1.8 mM CaCl2, 5 mM HEPES at pH 7.0) for 20 min to remove the surrounding follicular and epithelial cell layers. Oocytes were maintained at 18ºC in sterile unsupplemented modified Barth's medium containing (mM): 88 NaCl, 0.2 KCl, 2.4 $NaHCO_3$, 0.33 $Ca(NO3)_2$, 0.41 $CaCl_2$, 0.82 $MgSO_4$, 0.88 $KH_2PO_4$, 2.7 $Na_2HPO_4$, with gentamicin 70 µg ml−1 and adjusted to pH 7.4. Membrane currents were recorded with a standard two-electrode voltage clamp (Warner Instruments, Oocyte Clamp OC-725C) and digitized in a PC (Digidata 1200 and Axoscope 8.0 software, Axon Instruments). Oocytes were continually superfused with Ringer's solution (115 NaCl, 2 KCl, 1.8 $CaCl_2$, 5 Hepes) at room temperature (22ºC). The membrane was usually voltage clamped at -60 mV. Fetal Bovine Serum (FBS) (Sigma-Aldrich) diluted 1:1000 Ringer's solution was used for oocytes perfusion to achieve the generation of currents oscillations. Three different pH conditions were considered, Ringer's solution at pH 5.0, 7.0 and 9.0.

**2. Shannon Entropy**

Time series of calcium concentration flows, correlated to electrical potential measures, were first stationarized by performing a sliding window process as follows:

$$x_i \to x_i - \frac{x_{i-5} + x_{i-4} + x_{i-3} + x_{i-2} + x_{i-1} + x_i + x_{i+1} + x_{i+2} + x_{i+3} + x_{i+4} + x_{i+5}}{11}. \quad (1)$$

Then, the series were normalized within the (0,1) range. We will refer to this modified series as "stationarized series".

Let $\{p(x)\}_{x \in X}$ be a probability distribution over a finite set $X$ with states $x \in X$. The Shannon Entropy of $X$, denoted $H(X)$, is defined as

$$H(X) \equiv - \sum p(x) \cdot \log p(x). \qquad (2)$$

When the log is computed in base 2 (the case considered here), the Entropy is measured in bits. In particular, $H(X)$ is the expected value of the number of bits required to specify a concrete event in $X$ [70]. Thus, because $0 \leq p(x) \leq 1$, Entropy satisfies that $H(x) \geq 0$. The joint Entropy $H(X,Y)$ between two random variables $X$ and $Y$ is just an extension of (2) to 2-dimensions, i.e.

$$H(X,Y) \equiv - \sum p(x,y) \cdot \log p(x,y), \qquad (3)$$

where $\{p(x,y)\}_{x \in X, y \in Y}$ is the joint probability distribution of $X$ and $Y$ for events in which the variable $X$ is in the state x and simultaneously the variable $Y$ is in y.

### 3. Conditional Entropy

To measure how much uncertainty in future events is reduced by conditioning from past events, we calculated the Conditional Entropy. To compute it, we first denoted $X^P$ as the past of $X$ and $X^F$ as the future of $X$. The Conditional Entropy is then defined as

$$H(X^F | X^P) \equiv H(X^F, X^P) - H(X^P), \qquad (4)$$

being $H(X^P)$ the Shannon Entropy of $X^P$ and $H(X^F, X^P)$ the joint Entropy of $X^P$ and $X^F$ [70]. Thus, the Conditional Entropy is by definition the remaining uncertainty in $X^F$ known $X^P$. $H(X^F | X^P)$ is a positive-definite quantity such that its minimum value corresponds to $H(X^F | X^P) = 0$, which happens when knowing $X^P$ implies that the uncertainty in $X^F$ is completely determined. The maximum value corresponds to $H(X^F | X^P) = H(X^F)$, which occurs when knowing $X^P$ does not affect the uncertainty in $X^F$, so that both $X^P$ and $X^F$ are statistically independent variables.

### 4. Information Retention

We introduced the Information Retention index ($\eta$) as the Conditional Entropy normalized between 0 and 1, which allows comparing different time series within a common scenario. In particular, we defined it as

$$\eta \equiv 1 - \frac{H(X^F | X^P)}{H(X^F)}. \qquad (5)$$

Thus, this statistic will be closer to 0 as the future becomes harder to predict, and closer to 1 as the future becomes more dependent on the past.

# 5. Hurst exponent

The Scaled Windowed Variance Analysis is one of the most reliable methods that have been thoroughly tested on fBm signals [71]. In particular, we have used the bridge detrended Scaled Windowed Variance analysis (bdSWV) for the study of these temporal sequences of metabolic activities [72]. This method generates an estimation of the Hurst exponent (*H*) for each series. For a random process with independent increments, the expected value of *H* is 0.5. When *H* differs from 0.5, it indicates the existence of long-term memory, which is to say, dependence among the values of the process. If $H > 0.5$, it was produced by a biased random process which exhibits persistent behaviour. In this case, for several previous transitions, an increment on the phase-shift average value implies an increasing trend in the future. Conversely, a previously decreasing trend for a sequence of transitions usually implies a decrease for a similar sequence. Antipersistent behavior is obtained for $0 < H < 0.5$, a previously decreasing trend implies a probable increasing trend in the future and an increase is usually followed by decreases [71-72].

According to the bdSWV method, if the signal is of the form $x_t$, where $t = 1, ..., N$, then the following steps are carried out for each one of the window sizes $n = 2, 4, ..., \frac{N}{2}, N$ (if *N* is not a power of 2, then n takes the values 2,4,...,2$^k$, where k is the integer part of $\log_2 N$):

1) Partition of the data points in $\frac{N}{n}$ adjacent non-overlapping windows $\{W_1, ..., W_{\frac{N}{n}}\}$ of size *n*, where $W_i = \{x_{(i-1)\cdot n+1}, ..., x_{i\cdot n}\}$. If *N* is not a power of 2 and *N* is not divisible by *n*, then the last remaining points are ignored for this value of *n*. For instance, if $N = 31$ and $n = 4$, the first 28 points are partitioned into seven windows.

2) Subtraction of the line between the first and last points of the points in the *n*-th window.

3) For each $i = 1, ..., \frac{N}{n}$, calculation of the standard deviation $SD_i$ of the points in each window, by using the formula

$$SD_i = \sqrt{\sum_{t=(i-1)\cdot n+1}^{i\cdot n} \frac{(x_t - \bar{x}_i)^2}{n-1}}, \qquad (6)$$

where $\bar{x}_i$ is the average in the window $W_i$.

4) Evaluation of the average $\overline{SD}$ of the $\frac{N}{n}$ standard deviations corresponding to equation (6).

5) Observation of the range of the window sizes $n$ over which the regression line of $\log(\overline{SD})$ versus $\log(n)$ gives a good fit (usually some initial and end pairs are excluded).

6) In this range, the slope of the regression line gives the estimation of the Hurst coefficient $H$.

We have used the program bdSWV, available on the web of the Fractal Analysis Programs of the National Simulation Resource [73].

## 6. Dispersion Analysis

The Dispersion Analysis program is designed for the analysis of fractional Gaussian noise (fGn) [74].

For different bins of length $J$, varying from one to $\frac{n}{2}$, where $n$ is the length of the series, the standard deviation $SD(J)$ of the series formed by the mean of the $J$ consecutive values of the original series $x(i)$ are considered (that is, $SD(J)$ is the standard deviation of the series $y_J(i)$, where

$$y_J(i) = \frac{x(i) + \cdots + x(i+J-1)}{J}). \qquad (7)$$

Now, the relation between $\log SD(J)$ and $\log(J)$ is approximately linear:

$$SD(J) = SD(1) \cdot J^{H-1}, \qquad (8)$$

with slope $H$-1, where $H$ is the Hurst coefficient and with $SD(1)$ the standard deviation of a single point.

Previously, we have studied the Hurst exponent in a biochemical pathway [75-77] described by a system of differential equations with delay [78], in metabolic networks [79], and in experimental rabbit brain electrical signals [80].

**Results**

In order to study the informational properties of calcium signals, we have obtained experimental $Ca^{2+}$ measurements from seven different *Xenopus laevis* oocytes, each of them under three different external pH stimuli, gathering 21 time series, in which 130.000 data points were considered. Figure 1 shows three representative experimental calcium flows under three different pH conditions, Ringer's solution at pH 5.0, 7.0 and 9.0 (acid, neutral and basic pH).

First, the data has been analyzed according to Shannon's information theory, calculating the entropy of the time series (Shannon Entropy). This statistic quantifies the mean of the amount of information necessary to reproduce each calcium signal which is also interpreted as a measure of the average of uncertainty in the series. The obtained entropy values of our calcium data are in the order of $0.725 \pm 0.11 \ (mean \pm SD)$ bits. All values are shown in Table 1. To have a comparative reference for the amount of information in our calcium series, we calculated the entropy of 100 fractional Gaussian noise series (fully unpredictable events by definition, thus, series that need a lot of information to be transmitted). The fGn series were normalized between (0,1) as in the calcium flows, and the obtained entropy values were $0.987 \pm 0.01 \ (mean \pm SD)$. This implies that next to 73% in average of the information needed to represent a random event is necessary to represent our calcium series.

After analyzing the degree of uncertainty in the $Ca^{2+}$ ion flows we were interested in knowing how much uncertainty in the future is conditioned by the past. For it, we calculated the Conditional Entropy (4) applied to each stationarized time series (considered the past) and the stationarized time series t steps/lags further (considered the future). In other words, we cyclically permuted the data in order to compare this "future" series with the non-permuted ones. Thus, when the Conditional Entropy is minimum, the uncertainty is small and therefore the future becomes easier to predict. We can observe that there is a local minimum every 0.005 seconds in the experimental $Ca^{2+}$ series. The maximum values of (4) are shown in Table 1, and these results suggest the presence of a type of memory in the calcium flows, in the sense that the information known in the past of the series decreases the uncertainty in the future.

To quantify the long-term memory, we have first determined whether the series is a fractional Gaussian noise (fGn) or a fractional Brownian motion (fBm); fGn and fBm can be distinguished by calculating the slope of the Power Spectral Density plot [72]. The signal is said to exhibit power law scaling if the relationship between its Fourier spectrum and the frequency is approximated asymptotically by $S(f) \approx S(f_0)/f^\beta$ for adequate constants $S(f_0)$ and $\beta$. If $-1 < \beta < 1$, then the signal corresponds to an fGn. When $1 < \beta < 3$, then the signal corresponds to a fBm.

The regression line was estimated for the pairs $(\log S(f), \log f)$, where $f$ is the frequency and $S(f)$ the absolute value of the Fourier transform, and the $\beta$ constant was taken to be the opposite of the $x$ coefficient in that regression line. The analysis of the data in Table 2 shows that all the calcium signals that we have studied present power law scaling with $\beta$ in the range $1.507 - 2.991$, which suggests that the series are fBm. In Figure 2 the Spectral Density plot of the n1, n2 and n3 series are represented. The opposite of the slope have the values of 2.02 (n1, pH=5.0), 1.97 (n2, pH=7.0) and 2.37 (n3, pH=9.0), which shows that the process is a fractional Brownian motion (fBm).

A number of tools for estimating the long-term correlations of a fBm time series are available. The Scaled Windowed Variance analysis is one of the most reliable methods that have been thoroughly tested on fBm signals [72]. In particular, we have used the bridge detrended Scaled Windowed Variance analysis (bdSWV) on these temporal sequences [72,81]. By using this method, the estimated Hurst exponent values were around $0.191 \pm 0.101\ (mean \pm SD)$, which implies long-range memory and antipersistence in all experimental calcium series. The obtained Hurst exponent values are included in Table 2; in these results we have also observed that after an anova test, Hurst exponents were significantly different for time series corresponding to pH=9.0 in comparison to pH=7.0 ($p - value = 10^{-5}$) and pH=5.0 ($p - vaule = 10^{-4}$), but no distinction was found between pH=7 and pH=5 ($p - value = 0.42$).

In order to estimate the significance of our results, we have performed a shuffling procedure. If the data exhibit persistency, there is a long-range memory effect in place and the ordering of the data is fundamental. However, when we scramble the data, the structure of the series will be destroyed and, therefore, a Hurst analysis will show a value of $H$ close to 0.5. In this way, we have generated 21.000 scrambled series (1.000 for each calcium series) with the same number of points as the original ones (130.000 points) by taking values at random from the original data set without replacement. Next we have estimated the Hurst exponent $H$ for each of these scrambled series applying Dispersion Analysis [74], because after the shuffling they became fractional Gaussian noise, therefore we must not apply bdSWV [72,74] (for more details see section 6, Material and Methods). The result of the Hurst analysis for the scrambled series is $Hurst\ exp. = 0.499 \pm 0.01 (mean \pm SD)$ which indicates that in all scrambled series the long-term memory structure was destroyed. The Wilcoxon rank-sum test performed to compare the Hurst exponents of original calcium series with the exponents of the shuffled data, gave a $p-value = 2.13 \cdot 10^{-15}$. Therefore, the shuffling procedure analysis has shown that our Hurst exponent analysis of the $Ca^{2+}$ ion series presents a high significance level.

In Figure 3a the regression lines of a bdSWV process applied to an example of calcium series (n13, experiment 5, pH5.0) and the regression line of a Dispersion Analysis applied to the same data after randomly permuting all its 130.000 points are shown. In Figure 3b, 100 Hurst exponent values from one shuffled time series versus the exponent obtained from the original time series are illustrated. It can be observed that when the series were shuffled, the long-term memory completely disappeared ($Hurst\ exp. = 0.498 \pm 0.01$) (mean ±SD). In this figure, for more graphic clarity, instead of representing the 21.000 obtained Hurst exponent values from the respective scrambled time series, we have only illustrated 100 Hurst exponent values.

Once the presence of long-term memory in calcium flows was corroborated, we quantified the Information Retention (5) at each time step (for more details see section 4, Material and Methods). This statistic represents the amount of uncertainty that is reduced by knowing the past in order to determine the future. In our analysis, we have found that the values of this "certainty" oscillate periodically in all the calcium signals. Three examples of these informational structures for pH=5, pH=7 and pH9 can be observed in Figure 4.

In addition, we have observed that the highest and lowest peaks of the oscillation fluctuate erratically, an example of these dynamics is shown in Figure 5. To quantify the level of stability of these peaks, we have calculated the Largest Lyapunov exponent, a statistic associated to the predictability level of a dynamic system, when it is positive it reflects the presence of deterministic chaos. We calculated this statistic on the highest and lowest peaks separately. The Largest Lyapunov exponents of the highest peaks are around $0.062 \pm 0.03$ ($mean \pm SD$, all values are shown in table 2) and the ones of the lowest peaks are around $0.111 \pm 0.09$ ($mean \pm SD$). We also performed a B-L test for chaos [82, 83] to corroborate the presence of chaotic dynamics, and found that all p-values were equal to 1, implying that we do not reject the presence of chaos at $\alpha = 0{,}005$ level of significance.

In addition, a comparison between the structure of the Information Retention obtained from our experimental data with a different dynamic behaviors, a fractional Gaussian noise, is illustrated in Figure S1. As expected, the IR of the fGn does not exhibit significant oscillatory structure (range of IR values between 0 and 0.018).

Finally, to verify whether the structure of the Information Retention observed could be achieved by chance, we shuffled each stationarized time series and recalculated their $\eta_{max}$ 1.000 times; thus, 21.000 scrambled series were used for the shuffling procedure. In Figure S2a we show the $\eta_{max}$ values of our data versus the values from 100 shuffled procedures of one time series (for more graphic clarity, instead of including the 1.000 obtained $\eta_{max}$ from the respective scrambled time series, we have only illustrated 100 $\eta_{max}$). The probability distribution of $\eta_{max}$ values from our original data and the probability distribution of the $\eta_{max}$ values from the 100 shuffled series are respectively illustrated in Figure S2b and S2c. The result of this Information Retention analysis for the scrambled series indicates that the values from the shuffled series ($2.98 \cdot 10^{-5} \pm 1.78 \cdot 10^{-5}$, $mean \pm SD$) are at least four orders of magnitude smaller than the non-shuffled ones, implying that the informational structures in all shuffled series were completely lost, and therefore, the IR structure in the original calcium flows could not be found by chance ($p - value = 2.1 \cdot 10^{-15}$ in the Wilcoxon rank-sum test). In conclusion, the Information Retention structure analysis of the experimental $Ca^{2+}$ ion series presents a high significance level.

## Discussion

Calcium is a key chemical element of the cell which participates in a wide variety of physiological processes and plays a crucial role as a universal messenger in cellular metabolism and signaling.

Here, in order to understand some of the informational properties of the intracellular calcium signals we have studied experimental $Ca^{2+}$ series of *Xenopus laevis* oocytes under different external pH stimulus (acid, neutral and basic pH) by means of information-theoretic approaches.

In short, the main conclusions of our study are the following:

I. Calcium series carry information.

The amount of information contained in the experimental time series has been analyzed by means of Shannon's Entropy, finding that the quantity of information needed to reproduce these calcium signals is high, close to the amount needed to reproduce a completely unpredictable event such as fractional Gaussian noise.

II. Calcium series exhibit long-term memory.

The amount of uncertainty that can be predicted in the future when the past is known has been measured in the $Ca^{2+}$ ion fluxes using conditional Entropy. Our analysis has shown that a lot of the information in the future is conditioned by the past in every experimental time series (up to 86.3% of the uncertainty in the future on average was conditioned by past information). This implies a high dependency between values and also suggests the presence of memory.

The studied experimental series strongly correspond to the fractional Brownian motion (see the Power Spectral Density plot analysis), and in order to quantify the level of long-term memory in the $Ca^{2+}$ ion flows, the data was studied by means of the bridge detrended Scaled Windowed Variance analysis (Hurst exponent analysis) finding values for the Hurst exponent $0.05 < H < 0.35$, indicative of long-term memory effects in all series. Values of $H < 0.5$ are interpreted as characteristic of 'trend-reversing' or antipersistence. The behavior of the time series tends to reverse itself, for instance, a decreasing trend in the past usually implies an increasing trend on average in the future and conversely, an increase in the past is likely to be followed, on average, by a decrease. The high reliability of our Hurst analysis was tested with a shuffling procedure in which we scrambled the data in the original calcium signals. The values of the Hurst exponent measured in the original series (which are strongly non-Gaussian) were significantly different from the ones obtained in the randomly shuffled series. In the Wilcoxon rank-sum test performed to compare the Hurst exponents of the calcium series with the exponents of the shuffled data, we obtained a $p-value = 2.13 \cdot 10^{-15}$. Therefore, the significance of the values of the Hurst exponent measured in the original calcium series is extremely high, practically impossible to be obtained by chance.

III. Calcium series are characterized by a highly organized informational structure.

The analysis of Information Retention shows an informational structure in which the memory (uncertainty dependence level in the future as a function of the known information in the past) is oscillating. In addition, the highest and lowest peaks of the oscillation fluctuate erratically. The instability level of these maximums and minimums of the informational structure was analyzed by means of the Largest Lyapunov exponent, and all the calculated values were positive, which is characteristic of deterministic chaotic behavior. This dynamics were corroborated by B-L method for detecting chaos [82,83] with a p-value of 1, rejecting the absence of chaos.

The existence of chaotic patterns and long-term correlation properties in the informational structure of the calcium series may constitute a biological advantage. Chaotic patterns exhibit sensitive dependence on initial conditions, i.e., a small change in the initial state will lead to large changes in posterior system states and the fluctuations of the chaotic patterns are conditioned by the degree of perturbation of the initial conditions. These changes show exponential divergence, provoking fast separations in the chaotic trajectories. For "slow dynamic systems" the typical time scale of the chaotic fluctuations is of the order of 1 µs [84, 85] and in "very fast chaotic systems" the characteristic time scale is of about 1 ns [86,87]. Furthermore, different studies have shown that chaos permits fast and highly efficiency transmission of information [86].

The existence of chaos and long-term correlations in calcium series seems to constitute a biological advantage by allowing fast, efficient and specific responses during the adaptation of the cellular systems to environmental perturbations.

Besides, the informational structure with memory properties in the calcium series seems to be related to the dynamic memory of the recently proposed theory of the Cellular Metabolic Structure (CMS in short) [88]. At a systemic level, cells display a CMS, which behaves as very complex decentralized information processing system with the capacity to store metabolic memory [88, 89]. According to this newly published framework, the CMS exhibits two essential dynamic informational mechanisms. The first one occurs at the self-organized metabolic networks level, in which Hopfield-like attractor dynamics regulate the enzymatic activities. These attractors have the capacity to store functional catalytic patterns that can be correctly recovered by specific input stimuli. Hopfield-like metabolic dynamics are stable and can be maintained as long-term biochemical memory. The second mechanism occurs at the post-translational modification level. Specific molecular information can be transferred from the functional dynamics of the metabolic networks to the enzymatic activity involved in covalent post-translational modification, so that determined functional memory can be reversibly embedded in multiple stable molecular marks [88].

Since the beginning of the neuronal network modeling of associative memory, the connectivity matrix in the Hopfield network was assumed to result from a long-term memory learning process, occurring over a much slower time scale than neuronal dynamics [90-95]. Therefore, it is well accepted that the attractors emerging in neuronal dynamics described by Hopfield networks are the result of a long-term memory process. Besides, extensive physiological recordings of neuronal processes have revealed the presence of long range correlations in plasticity dynamics for measured synaptic weights. For instance, long tails in the synaptic distribution of weights correspond to a short-term memory in neural dynamics [96].

These studies and others support the thesis that neuronal dynamics exhibit both long-term and short-term memory, and the same may happen with the CMS. In fact, long-term correlations (mimicking short-term memory in neuronal systems) have also been analyzed in different metabolic processes not belonging to the neuronal cell-type. One of the most studied is the calcium-activated potassium channels, which have been observed in Leydig cells [57, 97], kidney Vero cells [59] and human bronchial epithelial cells [60]. Other biochemical processes also present long-term correlation*s,* for example, the intracellular transport pathway of Chlamydomonas reinhardtii [61], the NADPH series of mouse liver cells [62], and the mitochondrial membrane potential of cardiomyocytes [63]. Moreover, different biochemical numerical studies on enzymatic pathways and dissipative metabolic networks also show long-term correlations [75-77].

We think that the long-term memory in the calcium series could correspond to the short term memory of the dynamics in complex metabolic networks involved in the regulation of intracellular calcium signals, but this issue requires further detailed research.

Here, essential aspects of the information contained in experimental calcium fluxes have been exhaustively analyzed. We have quantified this information in bits, finding a highly organized informational structure which exhibits deterministic chaotic behavior, long term memory and complex oscillations of the uncertainty reduction in the future based on past values.

Understanding the informational properties of intracellular calcium signals is one of the key elements to elucidate the physiological functional coupling of the cell and the integrative dynamics of cellular life.

# References

1. Berridge MJ, Bootman MD, Lipp P. Calcium – a life and death signal. Nature. 1998;395: 645–648. doi: 10.1038/27094.

2. Berridge MJ, Bootman MD, Roderick HL. Calcium signalling: dynamics, homeostasis and remodelling. Nature Reviews Molecular Cell Biology. 2003;4: 517–529. doi: 10.1038/nrm1155.

3. Vaughan H, Thornton SD, Newsholme EA. The effects of calcium ions on the activities of trehalase, hexokinase, phosphofructokinase, fructose diphosphatase and pyruvate kinase from various muscles. Biochem J. 1973;132(3): 527-535.

4. Goldie AH, Sanwal BD. Allosteric control by calcium and mechanism of desensitization of phosphoenolpyruvate carboxykinase of Escherichia coli. J Biol Chem. 1980;255(4): 1399-1405.

5. Sola-Penna M, Da Silva D, Coelho WS, Marinho-Carvalho MM, Zancan P. Regulation of mammalian muscle type 6-phosphofructo-1-kinase and its implication for the control of the metabolism. IUBMB Life. 2010;62(11): 791-796. doi: 10.1002/iub.393.

6. Robinson R. Structure of signaling enzyme reveals how calcium turns it on. PLoS Biol. 2010;8(7): e1000427. doi: 10.1371/journal.pbio.1000427

7. Bikle DD, Xie Z, Tu CL. Calcium regulation of keratinocyte differentiation. Expert Rev Endocrinol Metab. 2012;7(4): 461-472. doi: 10.1586/eem.12.34

8. Komuro H, Kumada T. Ca2+ transients control CNS neuronal migration. Cell Calcium. 2005;37: 387–393. doi: 10.1016/j.ceca.2005.01.006

9. Hepler PK. The role of calcium in cell division. Cell Calcium. 1994;16(4): 322-330. doi: 10.1016/0143-4160(94)90096-5

10. Orrenius S, Gogvadze V, Zhivotovsky B. Mitochondrial oxidative stress: implications for cell death. Annu Rev Pharmacol Toxicol. 2007;47: 143–183. doi: 10.1146/annurev.pharmtox.47.120505.105122

11. Steinhorst L, Kudla J. Calcium and reactive oxygen species rule the waves of signaling. Plant Physiol. 2013;163(2): 471-485. doi: 10.1104/pp.113.222950.

12. Lohmann C, Finski A, Bonhoeffer T. Local calcium transients regulate the spontaneous motility of dendritic filopodia. Nature Neuroscience. 2005;8: 305–312. doi: 10.1038/nn1406

13. Szalai G, Krishnamurthy R, Hajnoczky G. Apoptosis driven by IP3 linked mitochondrial calcium signals. EMBO J. 1999;18: 6349–6361. doi: 10.1093/emboj/18.22.6349

14. Jangi SM, Ruiz-Larrea MB, Nicolau-Galmés F, Andollo N, Arroyo-Berdugo Y, Ortega-Martínez I, et al. Terfenadine-induced apoptosis in human melanoma cells is mediated through Ca2+ homeostasis modulation and tyrosine kinase activity, independently of H1 histamine receptors. Carcinogenesis. 2008;29(3): 500–509. doi: 10.1093/carcin/bgm292

15. Nicolau-Galmés F, Asumendi A, Alonso-Tejerina E, Pérez-Yarza G, Jangi SM, Gardeazabal J, et al. Terfenadine induces apoptosis and autophagy in melanoma cells through ROS-dependent and -independent mechanisms. Apoptosis. 2011;16(12): 1253-1267. doi: 10.1007/s10495-011-0640-y.

16. Barclay JW, Morgan A, Burgoyne RD. Calcium-dependent regulation of exocytosis. Cell Calcium. 2005;38(3-4): 343-353. doi: 10.1016/j.ceca.2005.06.012

17. Spitzer NC, Olson E, Gu X. Spontaneous calcium transients regulate neuronal plasticity in developing neurons. J Neurobiol. 1995;26: 316–324.

18. Furukawa R, Maselli A, Thomson SA, Lim RW, Stokes JV, Fechheimer M. Calcium regulation of actin crosslinking is important for function of the actin cytoskeleton in Dictyostelium. J Cell Sci. 2003;116: 187-196. doi: 10.1242/jcs.00220

19. Gomes TM, Spitzer NC. In vivo regulation of axon extension and pathfinding by growth-cone calcium transients. Nature. 1999;397: 350–355. doi: 10.1038/16927

20. Redmond L, Gosh A. Regulation of dendritic development by calcium signalling. Cell Calcium. 2005;37: 411–416. doi: 10.1016/j.ceca.2005.01.009

21. Sudhof TC. The synaptic vesicle cycle. Annu Rev Neurosci. 2004;27: 509–547. doi: 10.1146/annurev.neuro.26.041002.131412

22. Fields RD, Eshete F, Stevens B, Itoh K. Action potential-dependent regulation of gene expression: temporal specificity in Ca2 +, cAMP-responsive element binding proteins, and mitogen-activated protein kinase signaling. J Neurosci. 1997;17(19): 7252–7266.

23. Hu Q, Deshpande S, Irani K, Ziegelstein RC. [Ca2 +](i) oscillation frequency regulates agonist-stimulated NF-kappaB transcriptional activity. J Biol Chem. 1999;27(4): 33995–33998. doi: 10.1074/jbc.274.48.33995

24. Fields RD, Lee PR, Cohen JE. Temporal integration of intracellular Ca2+ signaling networks in regulating gene expression by action potentials. Cell Calcium. 2005;37(5): 433-442. doi: 10.1016/j.ceca.2005.01.011

25. Zhang S-J, Zou M, Lu L, Lau D, Ditzel DAW, Delucinge-Vivier C, et al. Nuclear Calcium Signaling Controls Expression of a Large Gene Pool: Identification of a Gene Program for Acquired Neuroprotection Induced by Synaptic Activity. Plos Genet. 2009;5(8): e1000604. doi: 10.1371/journal.pgen.1000604

26. Nakao A, Takada Y, Mori Y. [Calcium channels regulate neuronal function, gene expression, and development]. Brain Nerve. 2011;63(7): 657-667.

27. Bengtson CP, Bading H. Nuclear calcium signaling. Adv Exp Med Biol. 2012;970: 377-405. doi: 10.1007/978-3-7091-0932-8_17.

28. Hayer SN, Bading H. Nuclear Calcium Signaling Induces Expression of the Synaptic Organizers Lrrtm1 and Lrrtm2. J Biol Chem. 2014;2015 290: 5523-5532. doi: 10.1074/jbc.M113.532010

29. Dupont G, Combettes L, Bird GS, Putney JW. Calcium oscillations. Cold Spring Harb Perspect Biol. 2011;3(3). pii: a004226. doi: 10.1101/cshperspect.a004226.

30. Dawson AP. Calcium signalling: how do IP3 receptors work? Curr Biol. 1997;7(9): R544-R547. doi: 10.1016/S0960-9822(06)00277-6

31. Krauss G. Biochemistry of Signal Transduction and Regulation. 4$^{th}$ edition. Silverthorn; 2007.

32. Brini M, Calì T, Ottolini D, Carafoli E. Intracellular calcium Hhomeostasis and signaling. In: Banci L, Eeditors. Metallomics and the cell. Metal ions in life sciences. Springer; 2013; 12. pp 120-154. doi: 10.1007/978-94-007-5561-1_5.

33. Dupont G. Modeling the intracellular organization of calcium signaling. Wiley Interdiscip Rev Syst Biol Med. 2014;6(3): 227-237. doi: 10.1002/wsbm.1261.

34. De la Fuente IM. Quantitative analysis of cellular metabolic dissipative, self organized structures. Int J Mol Sci. 2010;11(9): 3540-3599. doi: 10.3390/ijms11093540.

35. De la Fuente IM. Metabolic Dissipative Structures in Systems Biology of Metabolic and signaling Networks: Energy, Mass and Information Transfer In: Aon MA, Saks V, Schlattner U, Editors. Systems Biology of Metabolic and Signaling Networks. Springer; 2013. pp 179-211

36. Goldbeter A. Computational approaches to cellular rhythms. Nature. 2002;420(6912): 238-245. doi: 10.1038/nature01259.

37. Goldbeter A. Biological rhythms as temporal dissipative structures. In: Special volume in memory of Ilya Prigogine: Adv Chem Phys. 2007;135: 253-295. doi: 10.1002/9780470121917.ch8.

38. Nicolis G, Prigogine I. Self-organization in nonequilibrium systems: From dissipative structures to order through fluctuations. Wiley; 1977.

39. Ishii K, Hirose K, Iino M. Ca2+ shuttling between endoplasmic reticulum and mitochondria underlying Ca2+ oscillations. EMBO. 2006;7: 390–396. doi: 1038/sj.embor.7400620

40. Newman EA. Propagation of intercellular calcium waves in retinal astrocytes and Müller cells. J Neurosci. 2001;21(7): 2215-2223.

41. Bernardinelli Y, Magistretti PJ, Chatton JY. Astrocytes generate Na+-mediated metabolic waves. Proc Natl Acad Sci USA. 2004;101(41): 14937–14942. doi: 10.1073/pnas.0405315101

42. Slaby O, Lebiedz D. Oscillatory NAD(P)H waves and calcium oscillations in neutrophils? A modeling study of feasibility. Biophys J. 2009;96: 417–428. doi: 10.1016/j.bpj.2008.09.044

43. Goldbeter A, Dupont G, Berridge MJ. Minimal model for signal-induced Ca2+ oscillations and for their frequency encoding through protein phosphorylation. Proc Nat Acad Sci. 1990;87: 1461–1465

44. Thomas AP, Bird GS, Hajnóczky G, Robb-Gaspers LD, Putney JW. Spatial and temporal aspects of cellular calcium signalling. FASEB J. 1996;10: 1505–1517

45. Schuster S, Marhl M, Höfer T. Modelling of simple and complex calcium oscillations. From single-cell responses to intercellular signalling. Eur J Biochem. 2002;269: 1333-1355. doi: 10.1046/j.0014-2956.2001.02720.x

46. Dupont G, Combettes L. Modeling the effect of specific inositol 1,4,5-trisphosphate receptor isoforms on cellular $Ca^{2+}$ signals. Biol Cell. 2006;98: 171–182. doi: 10.1042/BC20050032

47. Ventura AC, Sneyd J. Calcium oscillations and waves generated by multiple release mechanisms in pancreatic acinar cells. Bull Math Biol. 2006;68: 2205–2231. doi: 10.1007/s11538-006-9101-0

48. Dupont G, Abou-Lovergne A, Combettes L. Stochastic aspects of oscillatory $Ca^{2+}$ dynamics in hepatocytes. Biophys J. 2008;95: 2193–2202. doi: 10.1529/biophysj.108.133777.

49. Wieder N, Fink R, von Wegner F. Exact stochastic simulation of a calcium microdomain reveals the impact of $Ca^{2+}$ fluctuations on $IP_3R$ gating. Biophys J. 2015;108(3): 557-567. doi: 10.1016/j.bpj.2014.11.3458.

50. Rückl M, Parker I, Marchant JS, Nagaiah C, Johenning FW, Rüdiger S. Modulation of elementary calcium release mediates a transition from puffs to waves in an IP3R cluster model. Plos Comput Biol. 2015;11(1): e1003965. doi: 10.1371/journal.pcbi.1003965

51. Berridge M. Inositol trisphosphate and calcium signalling. Nature. 1993;361(6410): 315–325

52. Larsen AZ, Kummer U. Information Processing in Calcium Signal Transduction. In: Falcke M, Malchow D, editors. Understanding Calcium Dynamics: Experiments and Theory (Lecture Notes in Physics). Berlin: Springer Verlag. 2003.

53. Larsen AZ, Olsen L, Kummer U. On the encoding and decoding of calcium signals in hepatocytes. Biophys Chem. 2004;107: 83-99. doi: 10.1016/j.bpc.2003.08.010

54. De Pittà M, Volman V, Levine H, Ben-Jacob E. Multimodal encoding in a simplified model of intracellular calcium signaling. Cogn Process. 2009;10 Suppl 1:S55-70. doi: 10.1007/s10339-008-0242-y.

55. Prank K, Gabbiani F, Brabant G. Coding efficiency and information rates in transmembrane signaling. Biosystems. 2000;55(1-3):15-22.

56. Nakano T, Liu J-Q. Information transfer through calcium signaling In: Schmid A, Goel S, Wang W, Beiu V, Carrara S, editors. Lecture Notes of the Institute for Computer Sciences, Social Informatics and Telecommunications Engineering. Springer; 2009. pp 29-33

57. Nogueira RA, Varanda WA, Liebovitch LS. Hurst analysis in the study of ion channel kinetics. Braz J Med Biol Res. 1995;28(4): 491-496.

58. Varanda WA, Liebovitch LS, Figueiroa JN, Nogueira RA. Hurst analysis applied to the study of single calcium-activated potassium channel kinetics. J Theor Biol. 2000;206(3): 343-353.

59. Kazachenko VN, Astashev ME, Grinevic AA. Multifractal analysis of K+ channel activity. Biochemistry. 2007;2: 169-175.

60. Wawrzkiewicz A, Pawelek K, Borys P, Dworakowska B, Grzywna ZJ. On the simple random-walk models of ion-channel gate dynamics reflecting long-term memory. Eur Biophys J. 2012;41(6): 505-526. doi: 10.1007/s00249-012-0806-8.

61. Ludington WB, Wemmer KA, Lechtreck KF, Witman GB, Marshall WF. Avalanche-like behavior in ciliary import. Proc Natl Acad Sci USA. 2013;110(10): 3925-3930. doi: 10.1073/pnas.1217354110.

62. Ramanujan VK, Biener G, Herman B. Scaling behavior in mitochondrial redox fluctuations. Biophys J. 2006;90: L70-L72. doi: 10.1529/biophysj.106.083501.

63. Aon MA, Roussel MR, Cortassa S, O'Rourke B, Murray DB, Beckmann M, et al. The scale-free dynamics of eukaryotic cells. Plos One. 2008;3(11):e3624. doi: 10.1371/journal.pone.0003624.

64. Pahle J, Green AK, Dixon CJ, Kummer U. Information transfer in signaling pathways: a study using coupled simulated and experimental data. BMC Bioinformatics. 2008;9: 139. doi: 10.1186/1471-2105-9-139.

65. Wibral M, Lizier JT, Vögler S, Priesemann V, Galuske R. Local active information storage as a tool to understand distributed neural information processing. Front Neuroinform. 8:1. doi: 10.3389/fninf.2014.00001

66. Wibral M, Pampu N, Priesemann V, Siebenhühner F, Seiwert H, Lindner M, et al. Measuring information-transfer delays. Plos One. 2013;8(2): e55809. doi: 10.1371/journal.pone.0055809.

67. Dascal N, Landau EM, Lass Y. Xenopus oocyte resting potential, muscarinic responses and the role of calcium and guanosine 3',5'-cyclic monophosphate. J Physiol. 1984 Jul;352: 551-574.

68. Dascal N, Gillo B, Lass Y. Role of calcium mobilization in mediation of acetylcholine-evoked chloride currents in *Xenopus laevis* oocytes. J Physiol. 1985;366: 299-313.

69. Tigyi G, Dyer D, Matute C, Miledi R. A serum factor that activates the phosphatidylinositol phosphate signaling system in *Xenopus* oocytes. Proc Natl Acad Sci USA. 1990;87(4): 1521-1525.

70. Thomas JA, Thomas JA. Elements of information theory. 2$^{nd}$ ed. Wiley; 2006

71. Cannon MJ, Percival DB, Caccia DC, Raymond GM, Bassingthwaighte JB. Evaluating scaled windowed variance methods for estimating the Hurst coefficient of time series. Physica A. 1997;241: 606-626. doi: 10.1016/S0378-4371(97)00252-5

72. Eke A, Hermán P, Bassingthwaighte JB, Raymond GM, Percival DB, Cannon M, et al. Physiological time series: distinguishing fractal noises from motions. Pflugers Arch. 2000;439(4):403-415.

73. Raimond G. Fractal analysis programs of the national simulation resource. 2013. Available: http://www.physiome.org/software/fractal/

74. Caccia DC, Percival DB, Cannon MJ, Raymond GM, Bassingthwaight JB. Analyzing exact fractal time series: evaluating dispersional analysis and rescaled range methods. Physica A. 1997;246: 609-632. doi: 10.1016/S0378-4371(97)00363-4

75. De la Fuente IM, Martínez L, Aguirregabiria JM, Veguillas J. R/S analysis in strange attractors. Fractals. 1998;6(2): 95-100. doi: 10.1142/S0218348X98000110

76. De la Fuente IM, Martínez L, Benitez N, Veguillas J, Aguirregabiria JM. Persistent behavior in a phase-shift sequence of periodical biochemical oscillations. Bull Mathemat Biol. 1998;60(4): 689-702. doi: 10.1006/bulm.1997.0036

77. De la Fuente IM, Martínez L, Aguirregabiria JM, Veguillas J, Iriarte M. Long-range correlations in the phase-shifts of numerical simulations of biochemical oscillations and in experimental cardiac rhythms. Journal of Biological Systems. 1999; 7(2):113-130. doi: 10.1142/S0218339099000103

78. De la Fuente IM, Martínez L, Veguillas J and Aguirregabiria J M. Coexistence of multiple periodic and chaotic regimes in biochemical oscillations. Acta Biotheoretica. 1998; 46:37-51. doi: 10.1023/A:1000899820111

79. De La Fuente IM, Benítez N, Santamaría A, Veguillas J, Aguirregabiria JM. Persistence in metabolic nets. Bull Mathemat Biol. 1999; 61:573–595.

80. De la Fuente IM, Pérez-Samartín A, Martínez L, García MA, Vera-López A. Long-range correlations in rabbit brain neural activity. Annals of Biomedical Engineering. 2006; 34: 295-299. doi: 10.1007/s10439-005-9026-z

81. Herman P, Eke A. Nonlinear analysis of blood cell flux fluctuations in the rat brain cortex during stepwise hypotension challenge. J Cereb Blood Flow Metab. 2006; 26(9): 1189-1197. doi: 10.1038/sj.jcbfm.9600266

82. Ben Saïda A, Litimi H. High level chaos in the exchange and index markets. Chaos, Solitons & Fractals. 2013;2054: 90-95. doi: 10.1016/j.chaos.2013.06.004

83. BenSaïda A. Noisy chaos in intraday financial data: Evidence from the American index. Applied Mathematics and Computation. 2014;226: 258-265.

84. Blakely NJ, Illing L, Gauthier JD. Controlling fast chaos in delay dynamical. Syst Phys Rev Lett. 2004;92: 193901.

85. Garfinkel A, Spano ML, Ditto WL, Weiss JN. Controlling cardiac chaos. Science. 1992;257: 1230–1235.

86. VanWiggeren GD, Roy R. Communication with chaotic lasers. Science. 1998;279: 1198–1200. doi: 10.1126/science.279.5354.1198

87. Dronov V, Hendrey MR, Antonsen TM, Ott E. Communication with a chaotic travelling wave tube microwave generator. Chaos. 2004;14(1);30-37 doi: 10.1063/1.1622352

88. De la Fuente IM. Elements of the cellular metabolic structure. Front Mol Biosci. 2015;2: 16. doi: 10.3389/fmolb.2015.00016

89. De la Fuente IM, Cortes JM, Pelta DA, Veguillas J. Attractor Metabolic Networks. Plos One. 2013; 8(3): e58284. doi: 10.1371/journal.pone.0058284

90. Willshaw DJ, Buneman OP, Longuet-Higgins HC. Non-holographic associative memory. Nature. 1969;222: 960-962. doi: 10.1038/222960a0.

91.Van Heerden PJ. Models for the brain. Nature. 1970;225: 177-178. doi: 10.1038/225177a0.

92. Amari S-I. Homogeneous nets of neuron-like elements. Biol Cybern. 1975;17: 211-220. doi: 10.1007/BF00339367.

93. Hopfield JJ. Neural networks and physical systems with emergent collective computational abilities. Proc Nat Acad Sci USA. 1982;79(8): 2554-2558. doi: 10.1073/pnas.79.8.2554.

94. Amit DJ. Modeling brain function. – the world of attractor neural networks. New York: Cambridge University Press; 1989.

95. Hertz J, Krogh A, Palmer RG. Introduction to the theory of neural computation. Boston: Addison-Wesley Longman Publishing Co; 1991

96. Barbour B, Brunel N, Hakim V, Nadal JP. What can we learn from synaptic weight distributions? Trends Neurosci. 2007;30(12): 622-629. doi: 10.1016/j.tins.2007.09.005.

97. Bandeira HT, Barbosa CT, De Oliveira RA, Aguiar JF, Nogueira RA. Chaotic model and memory in single calcium-activated potassium channel kinetics. Chaos. 2008;18(3): 033136. doi: 10.1063/1.2944980.93. Hopfield JJ. Neural networks and physical systems with emergent collective computational abilities. Proc. Nat. Acad. Sci. USA. 1982;79(8): 2554-2558. doi: 10.1073/pnas.79.8.2554.

94. Amit D. Modeling Brain Function. New York: Cambridge University Press; 1989.

95. Hertz J, Krogh A, Palmer RG. Introduction to the theory of neural computation. Boston: Addison-Wesley Longman Publishing Co; 1991

96. Barbour B, Brunel N, Hakim V, Nadal JP. What can we learn from synaptic weight distributions?. Trends Neurosci. 2007;30(12): 622-629. doi:10.1016/j.tins.2007.09.005.

97. Bandeira HT, Barbosa CT, De Oliveira RA, Aguiar JF, Nogueira RA. Chaotic model and memory in single calcium-activated potassium channel kinetics. Chaos. 2008;18(3): 033136. doi: 10.1063/1.2944980.

# Figure 1

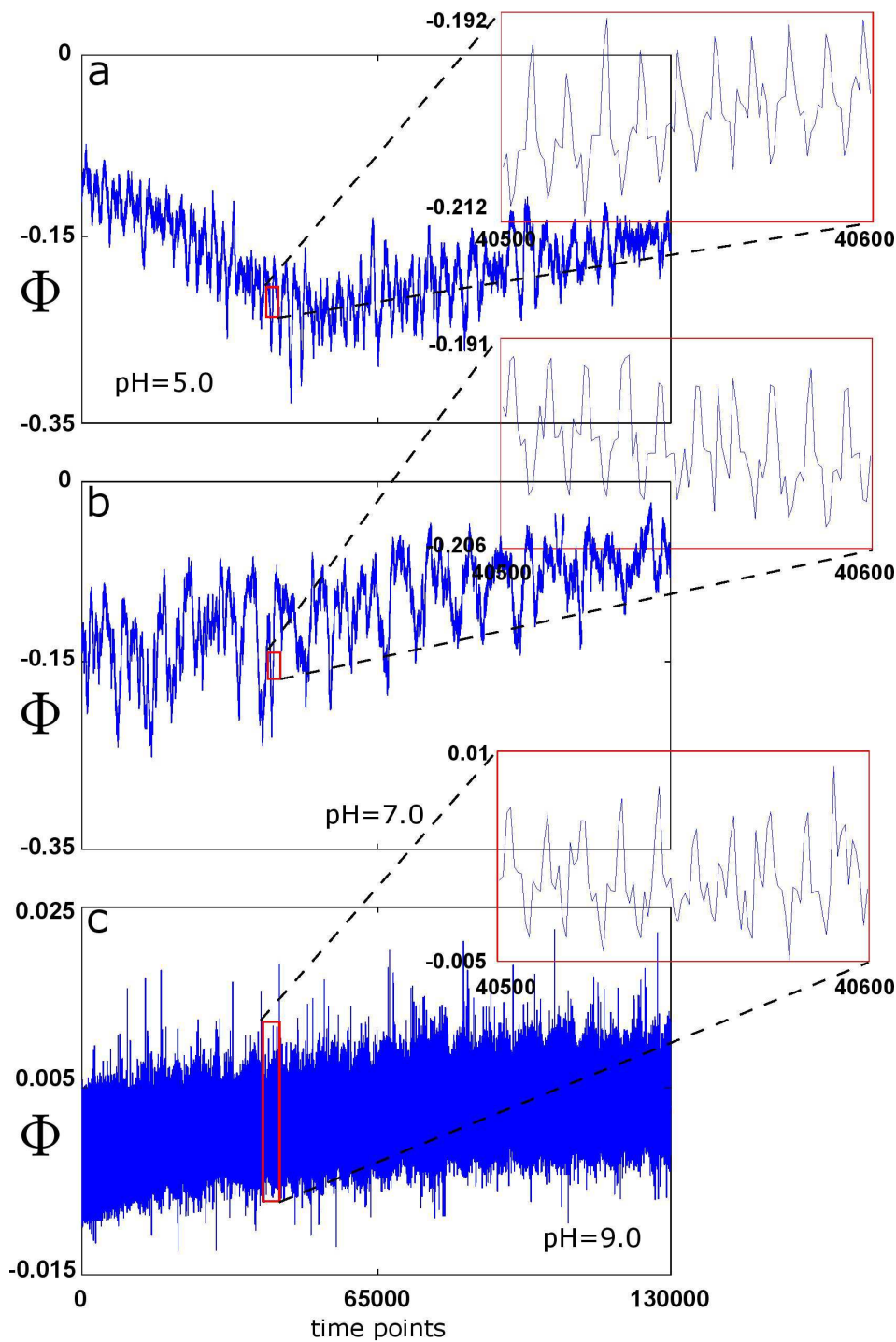

**Figure 1. Experimental calcium oscillations in *Xenopus laevis* oocyte.**
The calcium time series have been analyzed under three different pH conditions, Ringer's solution at pH 5.0, 7.0 and 9.0 (acid, neutral and basic pH). In the figure three prototype calcium signals obtained from the same cell are represented (experiment 4). a: pH 5.0 (n10), b: pH 7.0 (n11), c: pH 9.0 (n12). Each time series has 130.000 points, which correspond to an event 65.000 milliseconds long. The vertical axis ($\Phi$) correspond to the electrical potential measured in nanoampers (nA).

**Figure 2**

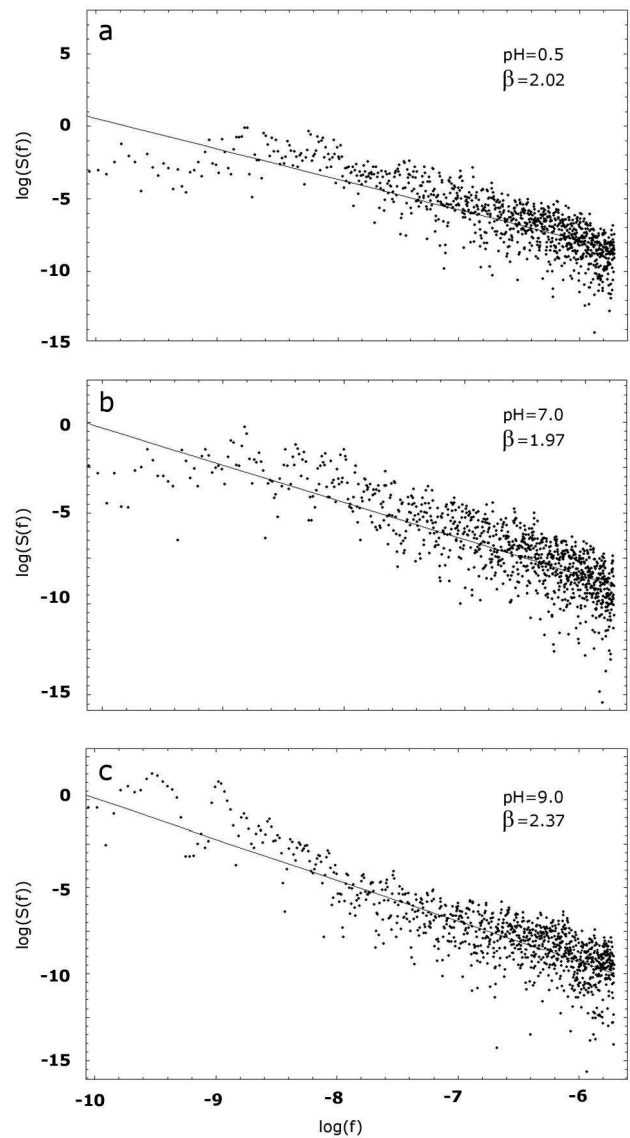

**Figure 2: Power spectrum estimation for the three pH values.**
Illustration of the power spectrum estimation (in a log-log scale) of the calcium series for different pH values. a: n1, experiment 1, pH=5.0; b: n2, experiment 1, pH=7.0; c: n3, experiment 1, pH=9.0. Slope values (from top to bottom) were 2.02, 1.97, 2.37, indicating for the three cases a fractional Brownian motion (fBm).

**Figure 3**

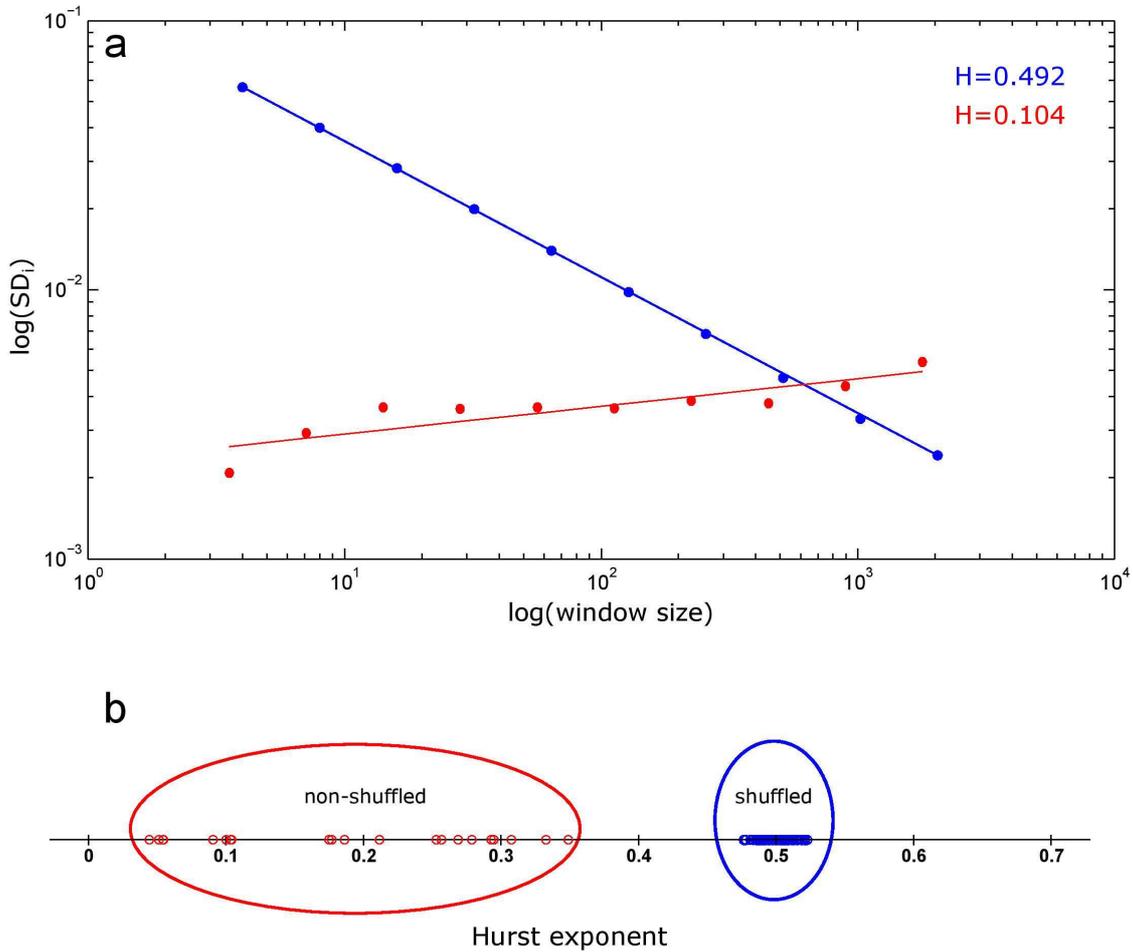

**Figure 3. Hurst exponents obtained by the bdSWV analysis in calcium series.**
a: In red, the slope of a log-log plot of the $SD_i$ versus the window size for a calcium series (n13, experiment 5, pH5.0) gives $H = 0.104$ for Hurst's exponents which indicates the presence of long term memory. In blue, the slope of a dispersion analysis applied to a 'scrambled' time series obtained by permuting randomly all of the 130.000 time points of the same series (n13). In these scrambled series the informational structure is destroyed and the values of $H$ are close to 0.5, which is indicative that the long-term memory has been lost. b: in red, Hurst exponent values of the all experimental time series; in blue, Hurst exponent values obtained after being shuffled all the original calcium series.

**Figure 4**

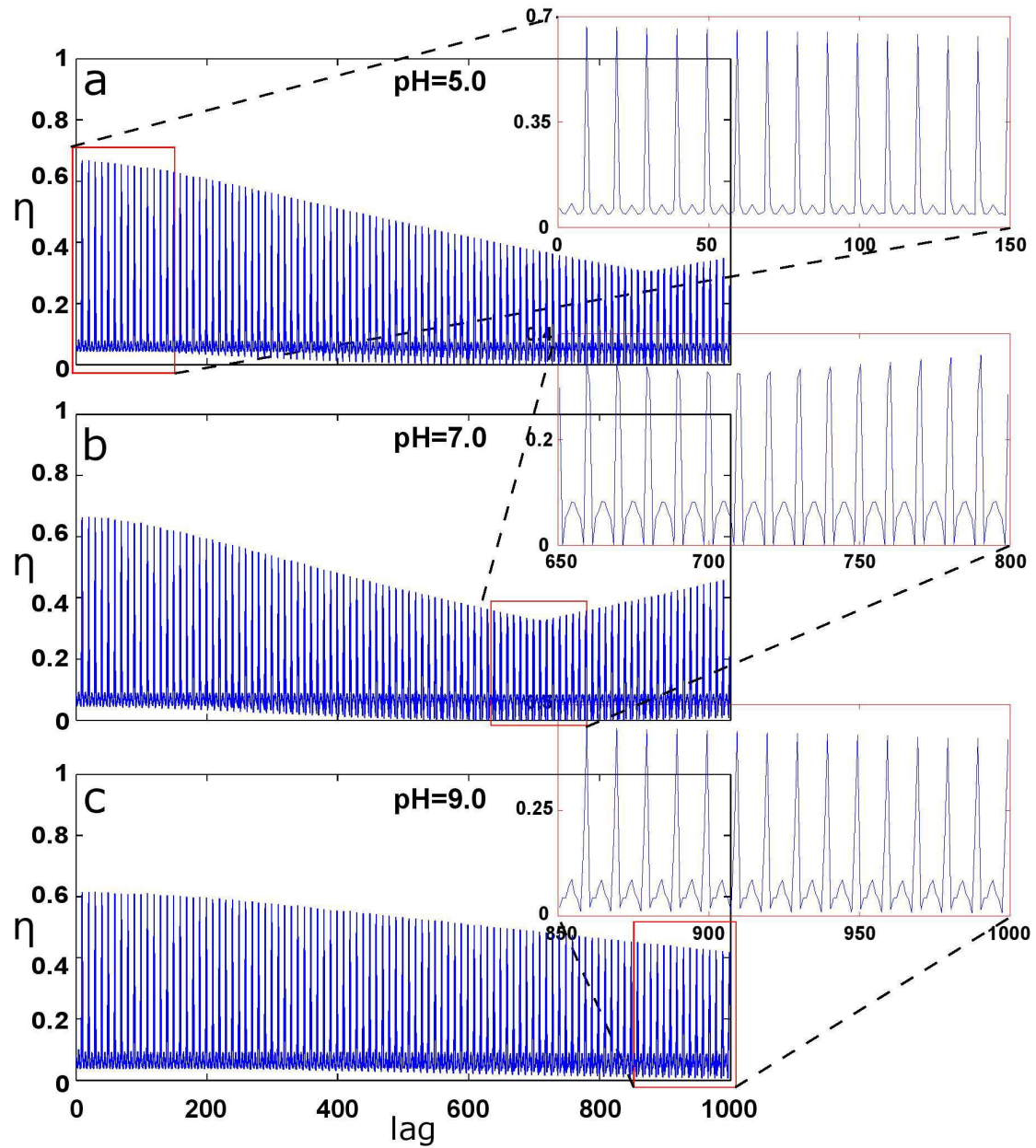

**Figure 4. Information Retention structure in calcium signals.**
Representation of the Information Retention calculated during 1000 lags under different pH stimuli (experiment 4), in which it can be observed that the uncertainty dependence level in the future as a function of the known information in the past is oscillating. a: IR from a stationarized calcium time series under pH5.0 (n10). b: IR from the same oocyte stimulated with pH7.0 (n11). c: IR from the same oocyte stimulated with pH9.0 (n12).

**Figure 5**

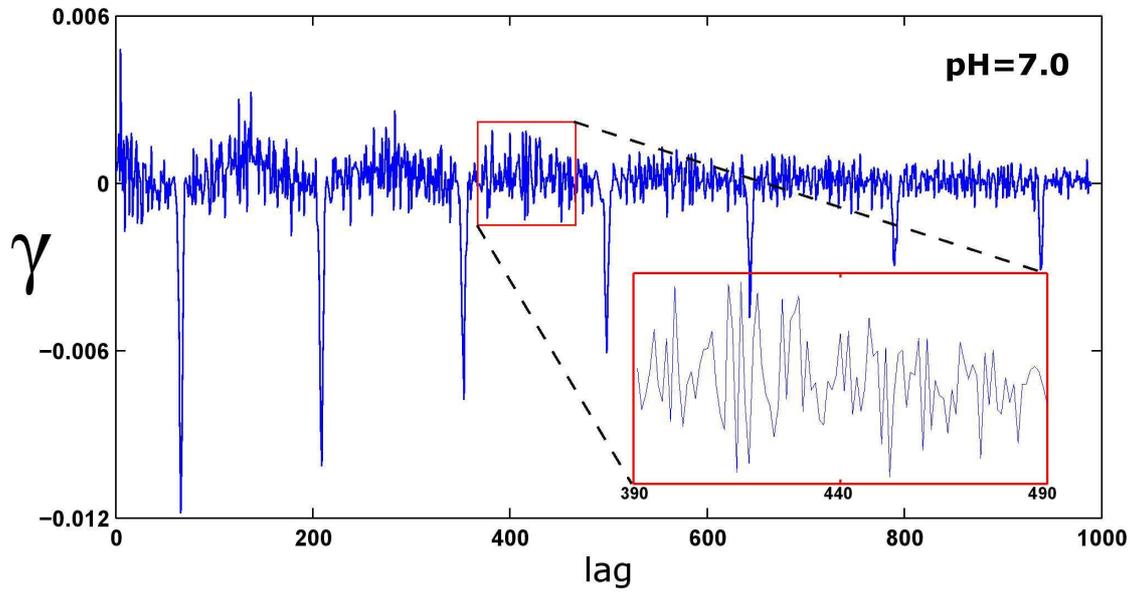

**Figure 5. Chaotic behavior emerges in the Information retention structure.**
Representation of the IR maximum values (n20, experiment 7, pH7.0). This maximums where obtained from the IR of a time series along 10.000 lags. The vertical axis correspond to the maximum Information Retention ($\gamma$), the horizontal axis shows the time lag.

**Figure S1**

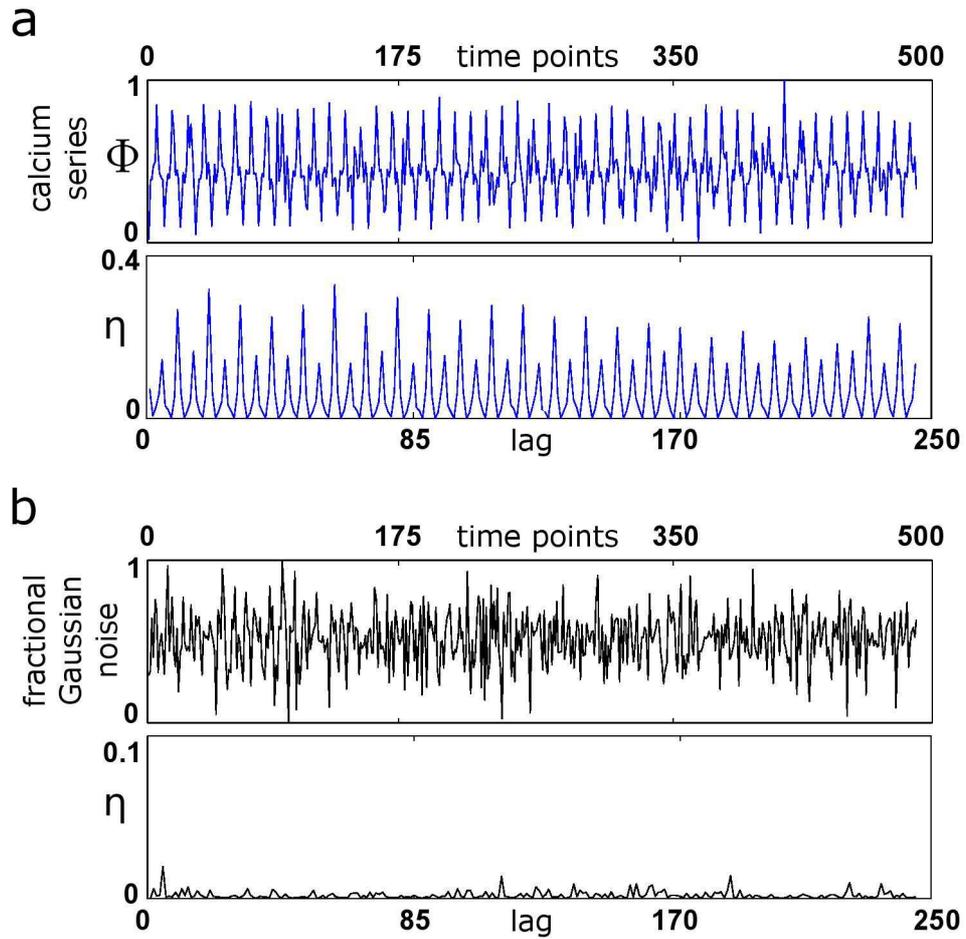

**Figure S1. Information Retention comparison between a calcium series and a fractional Gaussian noise.**
a: Top, a fragment of stationarized 500 points taken from an experimental time series (n18, experiment 6, pH9.0), the series is represented on the highest panel. Bottom, the Information Retention correspondent to this data. b: Stationarized Gaussian noise with mean $\bar{x} = 0$ and variance $\sigma = 1$, and its IR.

**Figure S2**

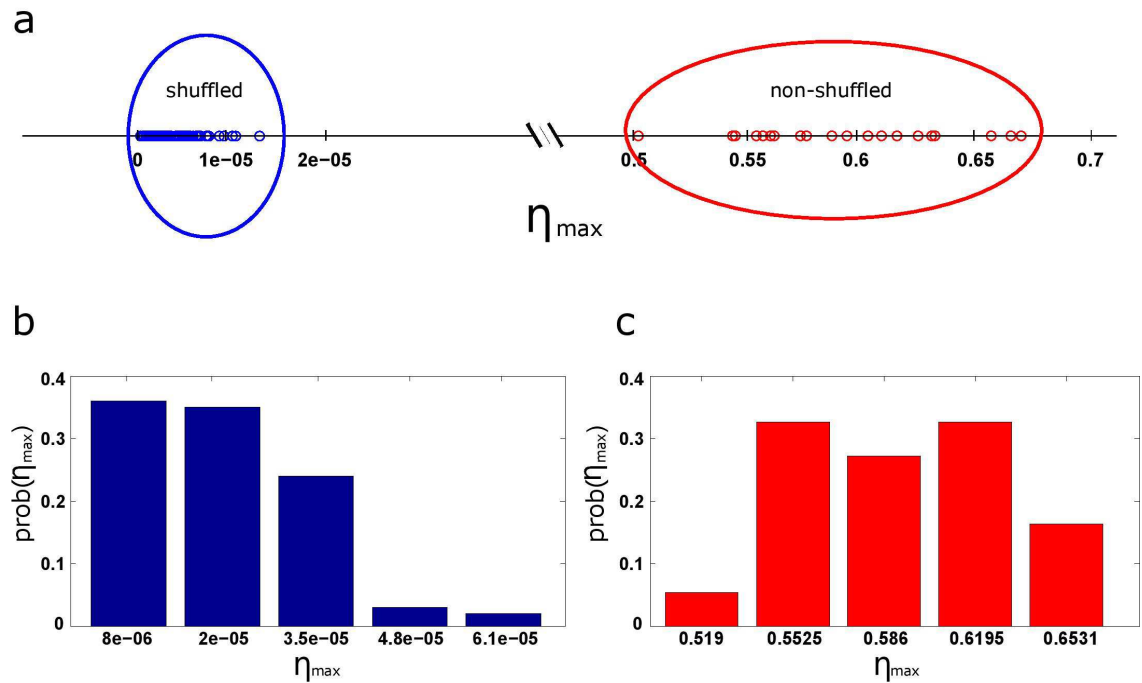

**Figure S2. Shuffling procedure analysis of the Information Retention.**
a: In red, maximum IR values belonging to the 21 experimental time series. In blue, the maximum IR values obtained after all the original calcium series were shuffled. The IR of the shuffled time series are four orders of magnitude smaller than the ones from the non-shuffled series, which indicates that in these scrambled series the information retention structure has been destroyed. b: Probability distribution of the IR from the shuffled time series. c: Probability distribution of the IR from the experimental time series.

**Table 1**

| Experiment | Stimulus | Number | *E* | *CE* | *η* |
|---|---|---|---|---|---|
| 1 | pH5.0 | n1 | 0,7903 | 0,3139 | 0,6028 |
|   | pH7.0 | n2 | 0,7838 | 0,2941 | 0,6247 |
|   | pH9.0 | n3 | 0,7107 | 0,2440 | 0,6567 |
| 2 | pH5.0 | n4 | 0,9156 | 0,4059 | 0,5566 |
|   | pH7.0 | n5 | 0,9174 | 0,4035 | 0,5601 |
|   | pH9.0 | n6 | 0,9095 | 0,3757 | 0,5869 |
| 3 | pH5.0 | n7 | 0,8410 | 0,3685 | 0,5617 |
|   | pH7.0 | n8 | 0,8314 | 0,3781 | 0,5447 |
|   | pH9.0 | n9 | 0,8133 | 0,3628 | 0,5538 |
| 4 | pH5.0 | n10 | 0,6921 | 0,2529 | 0,6320 |
|   | pH7.0 | n11 | 0,6101 | 0,2360 | 0,6086 |
|   | pH9.0 | n12 | 0,6561 | 0,2405 | 0,6307 |
| 5 | pH5.0 | n13 | 0,6110 | 0,2577 | 0,5731 |
|   | pH7.0 | n14 | 0,6274 | 0,2533 | 0,5935 |
|   | pH9.0 | n15 | 0,6670 | 0,2810 | 0,5757 |
| 6 | pH5.0 | n16 | 0,5525 | 0,2503 | 0,5433 |
|   | pH7.0 | n17 | 0,6429 | 0,2548 | 0,6027 |
|   | pH9.0 | n18 | 0,5495 | 0,2733 | 0,5022 |
| 7 | pH5.0 | n19 | 0,6728 | 0,2221 | 0,6698 |
|   | pH7.0 | n20 | 0,7118 | 0,2382 | 0,6653 |
|   | pH9.0 | n21 | 0,7330 | 0,2818 | 0,6155 |

The first column shows the number of the experiment, each corresponding to a single oocyte. The second column contains the pH stimuli applied to each experiment. The third shows the number assigned to each obtained series. The rest of the values correspond to informational measures: Shannon Entropy (*E*), conditional Entropy (*CE*) and Information Retention (*η*).

**Table 2.**

| Experiment | Stimulus | Number | $\beta$ | $H$ | $\lambda$ |
|---|---|---|---|---|---|
| 1 | pH5.0 | n1 | 2.0248 | 0.2455±0.0012 | 0.0477 |
|   | pH7.0 | n2 | 1.9723 | 0.1744±0.0016 | 0.0477 |
|   | pH9.0 | n3 | 2.3749 | 0.0950±0.0019 | 0.0474 |
| 2 | pH5.0 | n4 | 1.8460 | 0.2501±0.0009 | 0.0475 |
|   | pH7.0 | n5 | 1.6835 | 0.3521±0.0007 | 0.0559 |
|   | pH9.0 | n6 | 1.5669 | 0.1076±0.0017 | 0.1992 |
| 3 | pH5.0 | n7 | 1.8741 | 0.1830±0.0012 | 0.0477 |
|   | pH7.0 | n8 | 1.5075 | 0.2681±0.0010 | 0.0599 |
|   | pH9.0 | n9 | 1.7512 | 0.1071±0.0015 | 0.0520 |
| 4 | pH5.0 | n10 | 1.5834 | 0.2962±0.0012 | 0.0674 |
|   | pH7.0 | n11 | 1.6043 | 0.3174±0.0011 | 0.0730 |
|   | pH9.0 | n12 | 1.9086 | 0.0616±0.0023 | 0.0531 |
| 5 | pH5.0 | n13 | 2.0533 | 0.1040±0.0019 | 0.0662 |
|   | pH7.0 | n14 | 2.0738 | 0.1725±0.0016 | 0.0618 |
|   | pH9.0 | n15 | 1.8572 | 0.0589±0.0022 | 0.0477 |
| 6 | pH5.0 | n16 | 2.3889 | 0.3339±0.0009 | 0.0626 |
|   | pH7.0 | n17 | 2.4520 | 0.2949±0.0011 | 0.0524 |
|   | pH9.0 | n18 | 2.1698 | 0.0526±0.0022 | 0.0669 |
| 7 | pH5.0 | n19 | 2.8441 | 0.2174±0.0016 | 0.0473 |
|   | pH7.0 | n20 | 2.5817 | 0.2718±0.0013 | 0.0511 |
|   | pH9.0 | n21 | 2.9913 | 0.0621±0.0026 | 0.0474 |

The first column shows the number of the experiment, each corresponding to a single oocyte. The second column contains the pH stimuli applied to each experiment. The third shows the number assigned to each obtained series. The rest of the data corresponds to the values of Power Spectral Density slope ($\beta$), Hurst exponent ($H$) and Largest Lyapunov exponent ($\lambda$).